\documentclass[]{aastex7}

\usepackage{upgreek}

\submitjournal{ApJ}

\begin{document}

\title{Carbon-rich dust injected into the interstellar medium by Galactic WC binaries survives for hundreds of years}

\author[0000-0002-2806-9339]{Noel D. Richardson}
\affiliation{Department of Physics and Astronomy, Embry-Riddle Aeronautical University, 
3700 Willow Creek Rd, 
Prescott, AZ 86301, USA}
\email{noel.richardson@erau.edu}

\author[0009-0003-4158-4508]{Micaela Henson}
\affiliation{Department of Physics and Astronomy, Embry-Riddle Aeronautical University, 
3700 Willow Creek Rd, 
Prescott, AZ 86301, USA}
\email{SAMPSELM@my.erau.edu}

\author[0000-0002-4660-7452]{Emma P. Lieb}
\affiliation{Department of Physics and Astronomy, University of Denver,
2112 E. Wesley Ave.,
Denver, CO 80208, USA}
\email{emma.lieb@du.edu}

\author{Corey Kehl}
\affiliation{Department on Physics and Astronomy, Embry-Riddle Aeronautical University, 
3700 Willow Creek Rd, 
Prescott, AZ 86301, USA}
\email{KEHLC@my.erau.edu}

\author[0000-0003-0778-0321]{Ryan M. Lau}
\affiliation{NSF's NOIRLab, 950 N. Cherry Avenue, Tucson, Arizona 85719, USA}
\email{ryan.lau@noirlab.edu}

\author[0000-0002-8092-980X]{Peredur M. Williams}
\affiliation{Institute for Astronomy, University of Edinburgh, Royal Observatory, Edinburgh, EH9 3HJ, UK}
\email{pmw@roe.ac.uk}

\author[0000-0002-7762-3172]{Michael F. Corcoran}
\affiliation{CRESST II and X-ray Astrophysics Laboratory NASA/GSFC, Greenbelt, MD 20771, USA}
\affiliation{Institute for Astrophysics and Computational Sciences, The Catholic University of America, 620 Michigan Avenue, N.E. Washington, DC 20064, USA}
\email{corcoranm@cua.edu}


\author[0000-0002-7167-1819]{J. R. Callingham}
\affiliation{ASTRON, Netherlands Institute for Radio Astronomy, Oude Hoogeveensedijk 4, 7991 PD Dwingeloo, The Netherlands}
\affiliation{Anton Pannekoek Institute for Astronomy, University of Amsterdam, Science Park 904, 1098 XH Amsterdam, The Netherlands}
\email{callingham@astron.nl}

\author[0000-0002-1115-6559]{Andr\'e-Nicolas Chen\'e}
\affiliation{NSF's NOIRLab, 670 N. A`ohoku Place, Hilo, Hawai`i, 96720, USA}
\email{andre-nicolas.chene@noirlab.edu}

\author[0000-0002-6851-5380]{Theodore R. Gull}
\affiliation{Exoplanets \&\ Stellar Astrophysics Laboratory, NASA/Goddard Space Flight Center, Greenbelt, MD 20771, USA}
\email{tedgull@gmail.com}

\author[0000-0001-7515-2779]{Kenji Hamaguchi}
\affiliation{CRESST II and X-ray Astrophysics Laboratory NASA/GSFC,
Greenbelt, MD 20771, USA}
\affiliation{Department of Physics, University of Maryland, Baltimore County,
1000 Hilltop Circle, Baltimore, MD 21250, USA}
\email{kenjih@umbc.edu}

\author[0000-0002-2106-0403]{Yinuo Han}
\affiliation{Division of Geological and Planetary Sciences, California Institute of Technology,
1200 E. California Blvd., 
Pasadena, CA 91125, USA}
\email{yinuo@caltech.edu}

\author[0000-0001-9315-8437]{Matthew J. Hankins}
\affil{Arkansas Tech University, 215 West O Street, Russellville, AR 72801, USA}
\email{mhankins1@atu.edu}

\author[0000-0002-7648-9119]{Grant M. Hill}
\affiliation{W.M. Keck Observatory, 65-1120 Mamalahoa Highway, Kamuela, HI 96743, USA}
\email{ghill@keck.hawaii.edu}

\author[0000-0003-1495-2275]{Jennifer L. Hoffman}
\affiliation{Department of Physics and Astronomy, University of Denver,
2112 E. Wesley Ave.,
Denver, CO 80208, USA}
\email{jennifer.hoffman@du.edu}

\author[0000-0002-5449-6131]{Jonathan Mackey}
\affiliation{Astronomy \& Astrophysics Section, School of Cosmic
Physics, Dublin Institute for Advanced Studies, DIAS Dunsink
Observatory, Dublin D15 XR2R, Ireland}
\email{jmackey@cp.dias.ie}

\author[0000-0002-4333-9755]{Anthony F. J. Moffat}
\affiliation{D\'epartement de physique, Universit\'e de Montr\'eal, 1375 Avenue Th\'er\`ese-Lavoie-Roux, Montr\'eal (QC), H2V 0B3, Qu\'ebec, Canada}
\email{moffat@astro.umontreal.ca}

\author[0000-0003-2595-9114]{Benjamin J. S. Pope}
\affiliation{School of Mathematical \& Physical Sciences, 12 Wally's Walk, Macquarie University, NSW 2113, Australia}
\email{benjamin.pope@mq.edu.au}

\author[0000-0002-1131-3059]{Pragati Pradhan}
\affiliation{Department of Physics and Astronomy, Embry-Riddle Aeronautical University,
3700 Willow Creek Rd,
Prescott, AZ 86301, USA}
\email{pradhan@erau.edu}

\author[0000-0002-9213-0763]{Christopher M. P. Russell}
\affiliation{Department of Physics and Astronomy, Bartol Research Institute, University of Delaware, Newark, DE, 19716, USA}
\email{crussell@udel.edu}

\author[0000-0002-2090-9751]{Andreas A. C. Sander}
\affiliation{Zentrum f{\"u}r Astronomie der Universit{\"a}t Heidelberg,
Astronomisches Rechen-Institut,
M{\"o}nchhofstr. 12-14,
69120 Heidelberg, Germany}
\email{andreas.sander@uni-heidelberg.de}

\author[0000-0003-3890-3400]{Nicole St-Louis}
\affiliation{D\'epartement de physique, Universit\'e de Montr\'eal, 1375 Avenue Th\'er\`ese-Lavoie-Roux, Montr\'eal (QC), H2V 0B3, Qu\'ebec, Canada}
\email{stlouis@astro.umontreal.ca }

\author[0000-0001-7673-4340]{Ian R. Stevens}
\affiliation{School of Physics and Astronomy,
University of Birmingham, Birmingham, B15 2TT, UK}
\email{irs@star.sr.bham.ac.uk}

\author[0000-0001-7026-6291]{Peter Tuthill}
\affiliation{Sydney Institute for Astronomy, School of Physics, University of Sydney, NSW 2006, Australia}
\email{peter.tuthill@sydney.edu.au}

\author[0000-0001-9754-2233]{Gerd Weigelt}
\affiliation{Max Planck Institute for Radio Astronomy, 
Auf dem H\"ugel 69, 
53121 Bonn, Germany}
\email{weigelt@mpifr.de}

\author[0009-0006-7054-0880]{Ryan~M.T.~White}
\affiliation{School of Mathematical and Physical Sciences, Macquarie University, 
Sydney, 2113, NSW, Australia}
\email{ryan.white.astro@gmail.com}

\begin{abstract}

Some carbon-rich Wolf-Rayet stars (WC stars) show an infrared excess from dust emission. Dust forms in the collision of the WC wind with a companion star’s wind. As this dust is carried towards the ISM at close to the WCd wind speed and the binary continues through its orbit, a spiral structure forms around the system. The shape depends on the orbital eccentricity and period, as well as stellar parameters like mass-loss rates and terminal wind speeds. Imaging of the WCd binary WR 140 with \textit{JWST}/MIRI revealed 17 concentric dust shells surrounding the binary. We present new \textit{JWST} imaging of four additional WCd systems (WR 48a, WR 112, WR 125, and WR 137) that were imaged in 2024. In this analysis, we show that the dust is long-lived, detected with an age of at least 130 years, but more than 300 years in some systems. Longer duration measurements are limited by sensitivity. Regular spacing of dust features confirms the periodic nature of dust formation, consistent with a connection to binary motion. We use these images to estimate the proper motion of the dust, finding the dust to propagate out to the interstellar medium with motion comparable to the wind speed of the WC stars. In addition to these results, we observe unusual structures around WR 48a, which could represent dusty clumps shaped by photoevaporation and wind ablation like young proplyd objects. These results demonstrate that WC dust is indeed long-lived and should be accounted for in galactic dust budgets.

\end{abstract}


\section{Introduction} 

Classical Wolf-Rayet stars are evolved massive stars that are undergoing core helium fusion and have lost their hydrogen envelopes. These stars are relatively compact, have high effective temperatures, and drive fast, hot stellar winds. The typical mass-loss rates are on the order of $10^{-5}\ M_\odot\ {\rm yr}^{-1}$. 
The vast majority of the known classical WR stars are of the spectral types
WN or WC \citep{2001NewAR..45..135V, 2015MNRAS.447.2322R}, characterized by prominent nitrogen or carbon emission lines respectively, although a few so-called WO-type stars are known where the oxygen lines are more prominent than the carbon lines.

Most massive stars are found in multiple systems \citep{2012Sci...337..444S,2014ApJS..215...15S}, where binary interactions are seen to dominate the evolution, especially for shorter-period systems. Thus, it is reasonable to assume that many WR stars are formed through mass transfer or loss in binary systems. A good example is WR 140, an exceptional binary system consisting of a WC7pd star with an O5.5fc companion \citep{2011MNRAS.418....2F}. The orbit is extremely eccentric \citep[e=0.89;][]{2021MNRAS.504.5221T} with a period of 7.93 yr. \citet{2021MNRAS.504.5221T} found that the WR star likely was stripped by the O star companion, showing that even in long-period systems, binary interactions can be important. 

Dust around WC stars was first indicated by strong, often sudden variable infrared flux observed around several WC stars by \citet{1972A&A....20..333A}. WR 137 and WR 140 were seen to be variable in the infrared by \citet{1976ApJ...210..137H}, and, following further observations, interpreted as the rapid formation of dust by \citet{1977IAUC.3107....2W}, \citet{1978MNRAS.185..467W}, and \citet{1979MNRAS.187..183W}. Following these discoveries, more carbon-rich WR stars were found to have an infrared flux excess due to carbonaceous dust emission \citep[e.g.][]{1983A&A...118..301D}. The presence of dust around late-type WC stars is now indicated by a suffix of `d' being added to these stars' spectral type, which in turn relies both on spectral features and a spectral energy distribution. Time-series of the WCd stars show that two types of dust makers exist, both persistent dust creators that create dust all the time like WR 104, and episodic or periodic dust creators that show a burst of dust formation at certain times, often thought to happen near a periastron passage of an eccentric binary \citep[see ][for a discussion on their properties]{1995Ap&SS.224..267W}.

\citet{1999Natur.398..487T} demonstrated that the dust around the WC9 binary WR 104 was a nearly face-on spiral, indicative of a circular binary orbit with dust being formed in the shock cone. Further analysis of this system revealed a large spiral dust structure \citep{2008ApJ...675..698T}, while other studies quickly found large dust structures or spirals around other WCd stars \citep[e.g.][]{1999ApJ...525L..97M}.
Direct imaging with \textit{JWST} has dramatically enhanced our understanding of these structures. Prior to its launch, the best images of WR 140 taken in the mid-IR showed two concentric dust rings \citep{2009MNRAS.395.1749W}. The initial images of this system with MIRI showed at least 17 concentric rings \citep{2022NatAs...6.1308L}, which has been well modeled geometrically with a brief interval of dust formation near periastron passage every successive orbit \citep{2022Natur.610..269H}. A second epoch of imaging showed that the dust was propagating into the interstellar medium at a speed near the terminal wind speeds of the WC and O stars \citep{Lieb2025}. Here we present the first systematic mid-IR imaging survey with \textit{JWST} of dust structures around four other important dust-producing WC binaries. We introduce our sample in Section 2 and present the observational details in Section 3. In Section 4, we present the images and discuss the large-scale structures around these stars. We discuss our finding and show that in the environment surrounding WR 48a, we find structures that may be analogous to the Orion proplyds in Section 5. We conclude in Section 6 where we consider future goals with these and future data.

\section{The sample}

Our sample consists of four WCd stars that have been shown either to be in binary systems or to undergo periodic dust creation. Here we outline the observational history for each of these systems. The target binary parameters are given in Table \ref{tab:sample}. {blue} Some of our derived properties of the dust requires information on the distances to these systems so we have added discussion of the distance estimates for each binary here in addition to the known properties about the sample.

\begin{table}[]
    \centering
    \begin{tabular}{lcccccc}
    \hline 
    Target  &   Spec. &   Period  & Distance &  $v_\infty$ &  References (period) \\
     & ~Types  & (y) &  (kpc) & (km s$^{-1}$) &  \\
    \hline 
    WR 48a  &   WC8d+Oe         &   $\sim32$      &  $3.8\pm0.6$ & $1900\pm200$  & \citet{2012MNRAS.420.2526W} \\
    WR 112  &   WC8-9d+OB        &   $\sim 20$           &  $3.39	^{+0.89}	_{-0.84}$ &  1230$\pm$260  & \citet{2020ApJ...900..190L} \\
    WR 125  &   WC7d+O9III           &   28.12$\pm$0.1 & $5.88^{+0.72}_{-0.59}$  & 2700 & \citet{1994MNRAS.266..247W} \\
   & & & & &  \citet{2022ApJ...930..116E} \\
    WR 137  &   WC7pd+O9Ve           &   13.105$\pm$0.034        &  $2.10	^{+0.18}_{-0.16}$  & $1700\pm100$ & \citet{2020MNRAS.497.4448S} \\
        & & & & &  \citet{2024ApJ...977...78R} \\ \hline
    WR 140 & WC7pd + O5.5fc & 7.93$\pm0.001$ &  $1.64^{+0.08}_{-0.07}$  &  2860 & \citet{2011MNRAS.418....2F} \\
    & & & & &  \citet{2021MNRAS.504.5221T} \\ \hline 
    & & & & &  \citet{2022NatAs...6.1308L} \\ \hline 
    \end{tabular}
    \caption{WCd binaries in our sample and WR 140 for comparison. Listed references are for the spectral types and orbital period, respectively. The same reference is used for both in the cases of WR 48a and WR 112. Distances and terminal velocities are discussed in the text, with errors indicated when present in the referenced analyses. Periods are derived from IR photometry, RV orbits and/or estimates from IR geometric rotation.
    \label{tab:sample} }
\end{table}

\subsection{WR 48a}
WR 48a has a $\sim$32-year period/timescale for infrared outbursts and consists of a WC8 primary star with a companion that may be either an Oe star \citep{2012MNRAS.420.2526W} or a WN8h
\citep{2014MNRAS.445.1663Z} star, although several inconsistencies are present for a WN8h classification for the secondary (A.A.C. Sander, private communication). The wind has not been well-modeled to date, but the width of the \ion{C}{3} 5696 line is similar to that of WR 48-2 \citep{2014MNRAS.445.1663Z}, which shows a wind line width about $\sim 10-20\%$ greater than the same lines for WR 137 \citep{2012MNRAS.419.1871D}. WR 137 (Section 2.4) has a terminal wind speed of $1700\pm100$ km s$^{-1}$, so we adopt a terminal wind speed of $1900\pm200$ km s$^{-1}$ for our analysis, which is a bit higher than the average terminal velocity for WC8 stars observed by \citet{1994MNRAS.269.1082E}, which was 1620 km s$^{-1}$.
\citet{2012MNRAS.420.2526W} presented the long-term infrared light curve that was begun by \citet{1983A&A...118..301D}, \citet{1987A&A...182...91W}, and \citet{2003IAUS..212..115W}.
The infrared light curve shows both a long-period orbital modulation on the $\sim 32$ yr time-scale along with likely non-periodic mini-outbursts superimposed on the long-term variability. Previous imaging of the dust was done by \citet{2007ASPC..367..213M}, who used Gemini-South and TReCS at 12.7$\upmu$m in 2004 March, showed that the structure resembled a spiral structure similar to other WCd binaries. The infrared spectrum was measured with the \textit{ISO}-SWS instrument, and showed features at 7.7$\upmu$m and 6.4 $\upmu$m \citep{2003IAUS..212..115W, 2002ApJ...579L..91C}. The 6.4 $\upmu$m feature was shown to be due to carbonaceous dust around WR 137 by \citet{2023ApJ...956..109P}.

WR 48a is considered part of the nearby clusters Danks 1 (C 1310–624) and Danks 2 (C 1309–624) by \citet{1983A&A...118..301D, 1984A&A...132..301D}. The membership and association was strengthened by \citet{2004A&A...427..839C} who used mid-infrared data to show that there was triggered star formation in the vicinity of WR48a. \textit{XMM-Newton} observations revealed a highly luminous source at the location of WR 48a, which would make this system the most X-ray luminous WR binary in the Galaxy given the distance to the Danks 1 and 2 clusters \citep{2011ApJ...727L..17Z}. \citet{2014MNRAS.445.1663Z} also estimated that the mass-loss rate of the source to be a few$\times 10^{-4} M_\odot {\rm yr}^{-1}$ based on thermal radio emission and assuming an unclumped wind. The distance to WR 48a can be inferred either through the \textit{Gaia} parallax or from membership in the Danks 1 and Danks 2 clusters. The Galactic WR catalog\footnote{The catalog is maintained at https://pacrowther.staff.shef.ac.uk/WRcat/ and was described by \citet{2015MNRAS.447.2322R}.} lists the \textit{Gaia}-derived distance as $3.62^{+0.64}_{-0.49}$ kpc based on the analysis of \citet{2023MNRAS.521..585C}, which agrees well with the cluster-derived distance of $3.8 \pm 0.6$ kpc \citep{2012MNRAS.419.1871D}. For our analysis, we adopt the cluster distance.

\subsection{WR 112}

WR 112 was revealed to have dust shells by \citet{2002ApJ...565L..59M}, with \citet{2020ApJ...900..190L} finding the system to have a $\sim 20$-year orbital period from measured proper motion of its dust shell in infrared imaging. The companion star has not yet been clearly detected, although recent infrared spectroscopy is promising \citep{2025AAS...24535203B}. \citet{2020ApJ...900..190L} also presented an optical spectrum that confirmed a WC8 spectral type and a terminal wind speed of 1230$\pm$260 km s$^{-1}$. The source has been imaged in the infrared at multiple epochs between 2001 and 2019. These observations showed dust rings spaced with separations of $\sim$1.5\arcsec\ and exhibiting proper motion over these epochs. \citet{2020ApJ...900..190L} also found that the circumstellar dust around WR 112 is consistent with a nearly edge-on spiral with a wide opening angle, indicating that the OB companion should have a fairly high mass-loss rate. The edge-on geometry of WR 112 reconciled previous observations of highly variable non-thermal radio emission \citep[e.g.,][]{2002ApJ...566..399M,2007ApJ...655.1033M} that were inconsistent with a previously suggested face-on geometry, but were consistent with an edge-on system \citep{1993ApJ...402..271E, 2003A&A...409..217D}. \citet{2020ApJ...900..190L} also were able to measure dust masses and the dust creation rate of $2.6^{+1.0}_{-1.3} \times 10^{-6} M_\odot {\rm yr}^{-1}$ for this system, which is remarkably larger than that of the entire population of AGB and RSG stars in the LMC \citep[$\sim 9 \times 10^{-7} M_\odot {\rm yr}^{-1}$ as measured by ][]{2012ApJ...748...40B}. X-ray observations from \textit{Chandra} indicate an X-ray luminosity of about 10$^{33}$ erg s$^{-1}$ (Monnier et al., in prep) The \textit{Gaia} distance to WR 112 is not reliable for several reasons. \citet{2020MNRAS.493.1512R} flagged this target for its negative parallax and large astrometric noise in \textit{Gaia} DR2, and the EDR3 errors are extremely large ($> 3$kpc) with an expected distance of about 8 kpc \citep{2021AJ....161..147B}. The study by \citet{2020ApJ...900..190L} derived a distance of $3.39	^{+0.89}	_{-0.84}$ kpc which we use in our analysis.

\begin{figure}[t]
        \centering
        \includegraphics[width=1\linewidth,trim={0.25cm 0 0 0},clip]{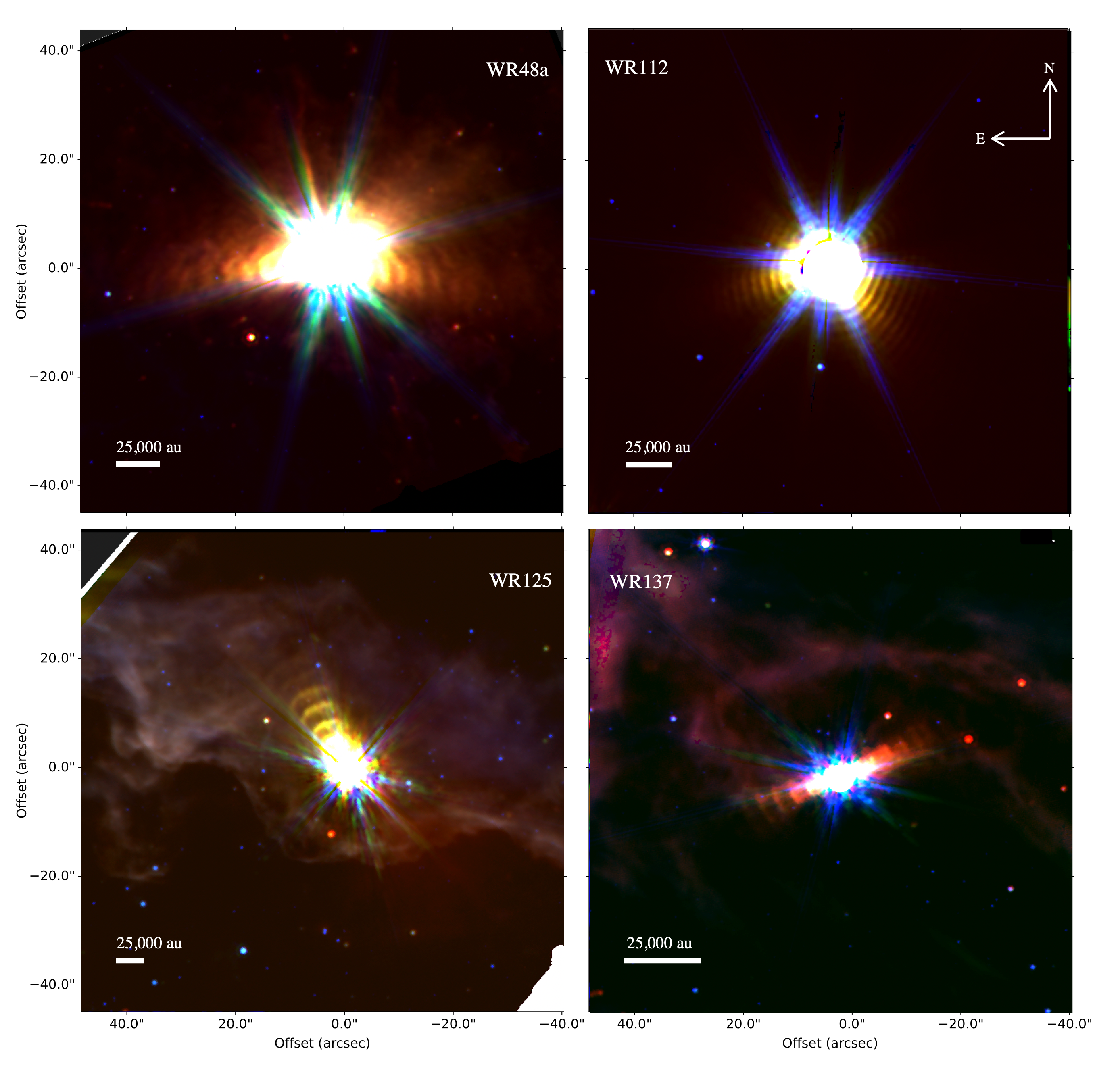}
    \caption{Data used for this study presented in a color-representation with the 7.7$\upmu$m filter in blue, the 15$\upmu$m filter in green, and the 21$\upmu$m filter in red and a white circle overlaid on the saturated cores. The orientation of each image is north to the top and east to the left, with coordinates centered on each star. For each image, we include a scale bar representing 25,000 AU at the distance discussed in Section 2 and Table 1.
    }
    \label{fig:rgb}
\end{figure}

\subsection{WR 125}

WR 125 is a near spectroscopic twin to WR 140, consisting of a WC7 star and a O9III companion in a 28.12-yr orbit \citep{2024ApJ...969..140R}. Its wind parameters have not been subjected to a full non-LTE model, but the \ion{He}{1} 10830\AA\ line shows a terminal wind speed of $\sim2700$ km s$^{-1}$ \citep{1994MNRAS.269.1082E}, although no error was reported. This is similar to the terminal wind speed of WR 140 used recently by \citet{2022NatAs...6.1308L} of 2860 km s$^{-1}$ based on the spectroscopic analysis of \citet{1989MNRAS.240..445W}. The system's spectroscopic similarities to WR 140 prompted \citet{1992MNRAS.258..461W} to monitor the system with infrared photometry, and an outburst of dust creation was observed beginning around 1990, which lasted for several years \citep{1992MNRAS.258..461W, 1994MNRAS.266..247W}. The system was seen to brighten again in the infrared by the \textit{NEOWISE} mission \citep{2019MNRAS.488.1282W}, prompting many new observations as the system should be near periastron during the infrared outbursts. The initial X-ray observations from both \textit{Swift} and \textit{XMM-Newton} showed an X-ray luminosity  of 10$^{33}$ erg s$^{-1}$ \citep{2019MNRAS.484.2229M}, while ground-based infrared photometry helped define the orbital period  \citep{2022ApJ...930..116E, 2024ApJ...969..140R}.
The dust was observed spectroscopically both from the ground \citep{2022ApJ...930..116E} and with airborne observations on SOFIA \citep{2024ApJ...969..140R}. \citet{2024ApJ...969..140R} used both archival and new optical spectroscopy to measure the first orbital elements of the system\footnote{The orbital elements were found with the assumption that the periastron passage happened at the mid-point between the start of dust creation beginning as observed with \textit{NEOWISE} \citep{2019MNRAS.488.1282W} and the peak of the $K-$band light curve \citep{2024ApJ...969..140R}}, finding a moderate eccentricity of 0.29$\pm$0.12 and an orbital period of 28.12 years based on a phase-dispersion minimization of the infrared light curve. WR 125 is the most distant object in our sample, with a \textit{Gaia} distance \citep{2023MNRAS.521..585C} of $5.88^{+0.72}_{-0.59}$ kpc that we use in our analysis. The \textit{Gaia} distance we use from \citet{2023MNRAS.521..585C} is comparable to the distance of  $6.58	^{+0.99}_{-	0.71}$ kpc from the statistical analysis of \citet{2021AJ....161..147B}.

\subsection{WR 137}

WR 137 was one of the first WR stars discovered \citep{1867CRAS...65..292W} and is spectroscopically similar to both WR 140 and WR 125, although the WC component has a slightly lower terminal wind speed than either of those systems. Its spectrum was modeled using the non-LTE code PoWR based on the spectrum across the ultraviolet through near-infrared, along with interferometric constraints on the fractional flux in the $H$-band by \citet{2016MNRAS.461.4115R}, who found the terminal wind speed to be $1700\pm100$ km s$^{-1}$. As a binary system, it holds the distinction of being one of only five WR binaries with a visually resolved orbit from interferometry \citep{2024ApJ...977...78R}. The eccentricity of the orbit is $0.3162\pm0.0023$, with the orbit nearly edge-on ($i = 97.138^\circ$). The dust plume, reconstructed from \textit{JWST}/NIRISS aperture masking interferometry \citep{2024ApJ...963..127L} is narrow and is in agreement with the orbital geometry from \citet{2024ApJ...977...78R}. The companion O9V star is an Oe star with a fairly stable disk \citep{2020MNRAS.497.4448S}. The geometry of the Oe disk with respect to the orbital plane may allow dust to condense through an interaction with the disk and outflowing WC wind, which likely partly explains why the infrared light curve is not as strictly periodic as in WR 140 \citep{2023ApJ...956..109P}. \citet{2023ApJ...956..109P} also found spectroscopic signatures of new dust formed during the most recent periastron using SOFIA observations. \citet{2024ApJ...977...78R} discussed the implications of different distances for the system, and we adopt their preferred distance of $2.10^{+0.18}_{-0.16}$ kpc.

\section{Observations and Data Reductions}
The imaging observations we present in this work were taken with \textit{JWST}'s Mid-Infrared Instrument (MIRI; \citealt{Wright2015}). The filters used for all sets of images were F770W, F1500W, and F2100W, which correspond to wavelengths of 7.7 $\upmu$m, 15 $\upmu$m, and 21 $\upmu$m, respectively; these IR filters are ideal for probing the thermal emission of the dust in these systems. These images were taken during \textit{JWST} Cycle 2 with program ID 4093 (PI: N. Richardson). Each set of observations employed one set of 4-point dithers to improve PSF sampling, reduce detector artifacts, and enhance contrast, and we tabulate the observation dates and total exposure times in Table \ref{tab:obs}. 

\begin{table}[t!]
    \centering
    \begin{tabular}{lcccc}
\hline
       Star	&	Date	& 	770W exposure	& 	1500W exposure	&	2100W exposure	\\	\hline
WR 48a	&	2024 July 16	& 	1110 s	& 	1110 s	&	4473 s	\\	
WR 112	&	2024 May 07	& 	222	s & 	111 s	&	1143 s	\\	
WR 125	&	2024 Aug 13	& 	1110 s	& 	2231 s	&	6715 s	\\	
WR 137	&	2024 Aug 05	& 	1110 s	& 	2231 s	&	4473 s	\\	\hline
    \end{tabular}
    \caption{Observation dates and exposure times.}
    \label{tab:obs}
\end{table}

The innermost shells are saturated in our images due to the extremely bright central cores. We reduced some of the effects of this saturation by running our own Stage 2 images from the uncalibrated data, changing the calibration parameters from the default pipeline to reduce the growth of saturated pixels. This step is crucial to the PSF subtraction steps described below for finding a precise center point for our targets in each dither frame image so that the center of the PSF model can be matched to the intersection point of the diffraction spikes through the source. We re-ran our own Stage 1 and Stage 2 reductions from the uncalibrated data, using {\tt calwebb\textunderscore detector1} and {\tt calwebb\textunderscore image2} from the {\tt jwst} pipeline version 1.16.0 and CRDS context 1293.  We updated the calibration parameters from the default pipeline to reduce the growth of saturated pixels by making the following changes: skipping the first frame step, setting the parameter {\tt suppress\textunderscore one\textunderscore group=False} for the {\tt ramp\textunderscore fit} step, and setting {\tt n\textunderscore pix\textunderscore grow\textunderscore sat=0} for the jump step. These changes allow the pipeline to make flux estimates even from pixels that saturated after only a single read at the start of the ramp, which was the case for the very bright inner regions of all the binaries in our sample. We found that the inner pixel measurements obtained in this way are sufficiently precise, and are preferable to leaving those pixels as missing values even though flux estimates like these are intrinsically less precise than ramp fits to unsaturated pixels. Once the new Stage 2 images were obtained, we suppressed contamination by the PSF diffraction spikes and performed a background subtraction. 

The MIRI PSF includes six bright diffraction spikes from the hexagonal primary mirror and two fainter spikes caused by the secondary mirror support. At wavelengths below $\sim10 \mu$m, an additional four faint perpendicular spikes (denoted as the ``cruciform'') arise from internal scattering within the MIRI detectors \citep{Gaspar2020,Dicken2024}. We used {\tt WebbPSF} \citep{Perrin2012} to create our PSF model for each Stage 2 image in our sample (4 dither frames per filter per target yielding 48 images in total). We aligned this model to the centroid of our target such that the diffraction spikes in the model match in pixel-space to the diffraction spikes in our images, and scaled the model to meet the flux of its corresponding science image. The PSF model was then subtracted off of each Stage 2 image. For more details on the PSF subtraction pipeline, including the background subtraction methods, please see \citet{Lieb2025}. Finally, we created Stage 3 mosaic images for each system using {\tt calwebb\textunderscore image3} from the {\tt jwst} pipeline.

\section{Large-scale spiral structures around the systems}

\begin{figure}[t]
    \centering
    \includegraphics[width=0.9\linewidth]{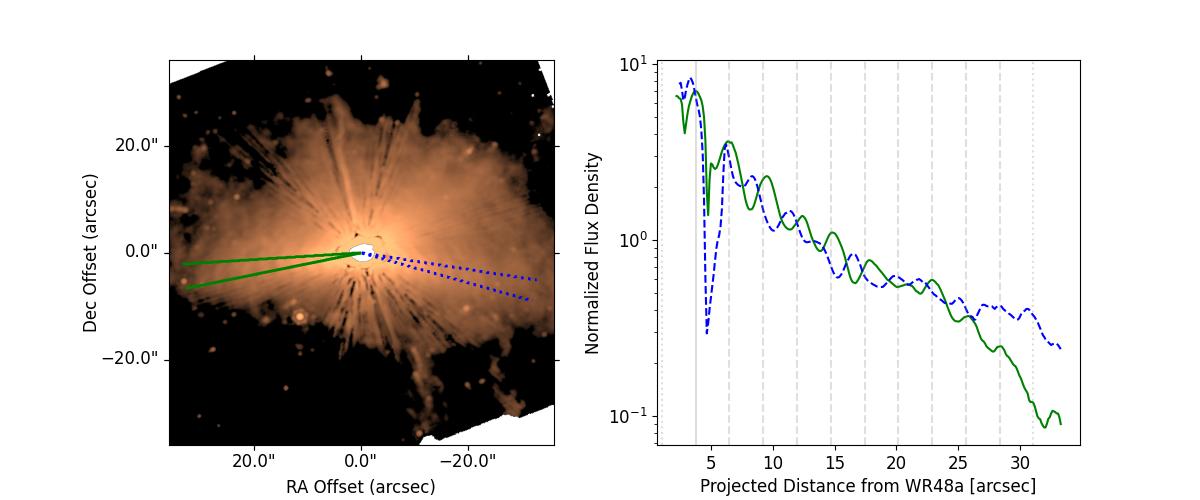}
    \includegraphics[width=0.9\linewidth]{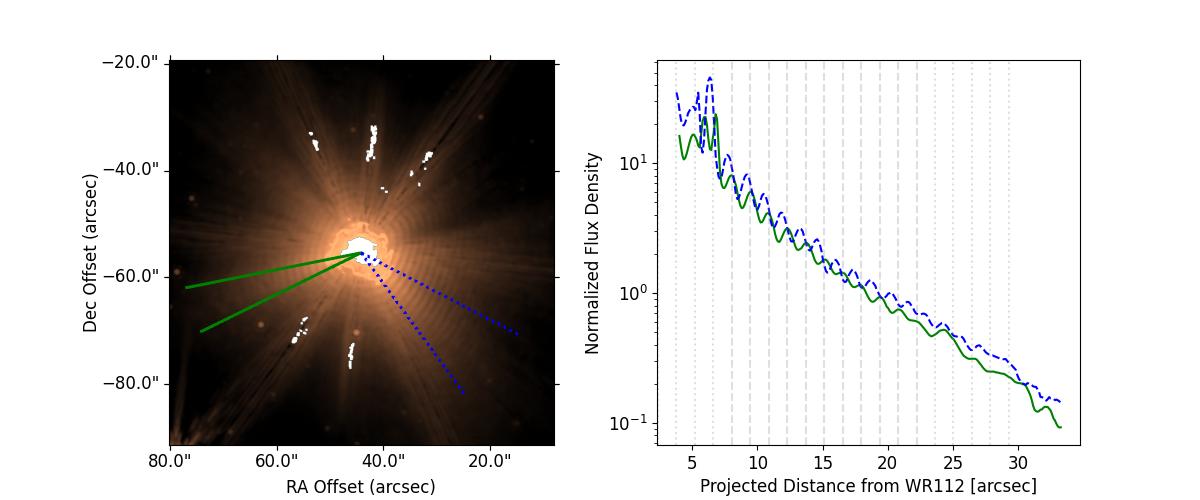}
    \caption{21$\upmu$m images of WR 48a (top) and WR 112 (bottom). In the data image, we overplot radial lines from the central binary outward which show the boundaries of the directions in which we integrated to give the radial plots shown on the right. The number of apparent rings around the WC star gives a minimum lifetime of the dust. We show the first well-defined ring as a vertical solid line, with the average distance to the next ring showed with each dashed vertical line. If the central star and dust saturated in the image, we include vertical dotted lines at those locations or at farther out locations where the measurement couldn't be directly made but peaks may be present. We also include the average peak separation in Table \ref{tab:lifetime} which are measured from an average of all directions used where the dust is readily detected. We normalized this flux distribution extracted to be $\sim 1$ near the center of the plot for the normalized flux density}. 
    \label{fig:radial1}
\end{figure}

In Fig.~\ref{fig:rgb}, we show the PSF-subtracted three-color images of our targets. In these systems, especially in WR 112, we see that the PSF-subtraction left some residual diffraction spikes, especially in the bluest F770W filter. In all cases, we see large, extended structures around the stars that can be described as repeated dust structures that are similar to those observed with \textit{JWST} around WR 140 \citep{2022NatAs...6.1308L, Lieb2025}.

In the cases of WR 48a, WR 112, and WR 137, we can compare our \textit{JWST} images to previously published images. WR 48a was imaged with Gemini-S and T-ReCs \citep{2007ASPC..367..213M}, showing a dust structure that is generally consistent with the nested dust rings seen in the \textit{JWST} image. Our images of WR 112 also exhibit similar geometry to the images from the ground \citep[e.g.][]{2020ApJ...900..190L}, which indicated the orbital period to be on the order of $\sim20$ yr. Finally, reconstructed images from aperture masking interferometry by the NIRISS instrument on \textit{JWST} only show the central dust arc \citep{2024ApJ...963..127L}, but these qualitatively agree with the extended emission seen with HST/NICMOS \citep{1999ApJ...522..433M}. The resolved dust plume from \textit{JWST}/NIRISS was consistent with dust formation happening in the orbital plane of the WC and Oe binary that was resolved by \citet{2024ApJ...977...78R}.

\begin{figure}[t]
    \centering
    \includegraphics[width=0.9\linewidth]{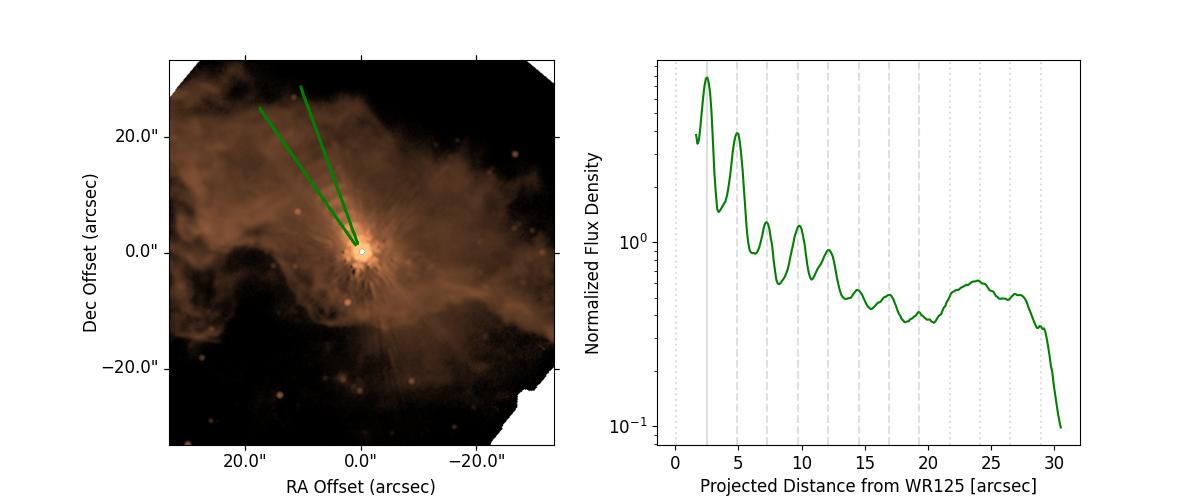}
    \includegraphics[width=0.9\linewidth]{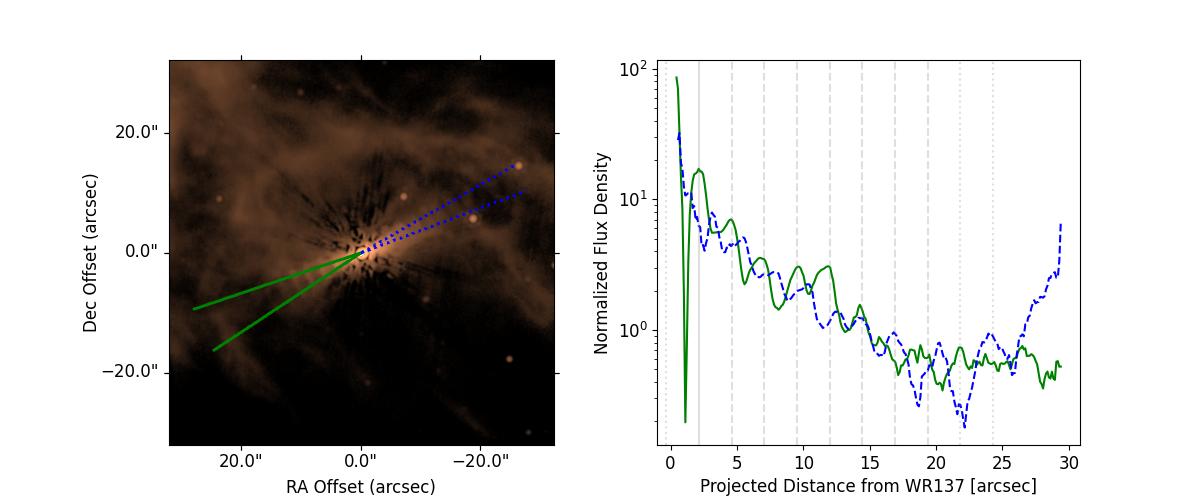}
    \caption{21$\upmu$m images of WR 125 (top) and WR 137 (bottom) in a similar format as to those of WR 48a and WR 112 in Fig.~\ref{fig:radial1}.  }
    \label{fig:radial2}
\end{figure}

To determine the lifetime of the dust clouds around the WCd stars, we extracted flux profiles in conical/radial directions from the binary. We show these in Figs.~\ref{fig:radial1} and \ref{fig:radial2}. From each image, we then counted the number of rings observed, and under the assumption that each ring corresponds to a single orbit of the binary, we were able to determine the extent of the furthermost dust ring and the time of ejection. These minimum ``dust detection times" are given in Table \ref{tab:lifetime}. For WR 125, we only extract in one direction as it only shows dust in a short range of azimuth. This is consistent with dust condensation occurring for only a relatively short portion of the binary orbit, $\sim 2.7$ yrs \citep{1994MNRAS.266..247W, 2024ApJ...969..140R}, of its 28.1-yr orbit.

In most of these radial plots, we see a general agreement in different azimuthal directions. In the case of WR 137, where we know the most about the dust formation in relation to its orbit, this is easiest to understand as we know that the dust forms in a fairly quick burst near periastron. Based on the infrared light curve recently presented by \citet{2023ApJ...956..109P} and the visual orbit from \citet{2024ApJ...977...78R}, we see that the plume starts to form as the stars are near a quadrature in their orbit, with dust production continuing until after the conjunction in the orbit. The result is that we see dust emanating in both directions along the nearly edge-on orbital plane. Conversely the dust spirals around WR 48a likely have an inclination closer to 60$^\circ$ from our line of sight, but the orbital period is much longer with continuous dust production \citep{2012MNRAS.420.2526W} across the orbit. This results in the radial plots showing maxima in one direction but a minimum at the same distance in the opposite direction as the stars move halfway around their orbits in between the times when the dust peaks form.

\begin{table}[]
    \centering
    \begin{tabular}{lcccccc}
  \hline
Star	&	$N_{\rm rings}$	&	$T_{\rm dust\ detected}$	&	Ring Spacing	&	Proper Motion & $v_{\rm dust}$ & $v_\infty$	\\
	&		&	(years)	&	(\arcsec)	&	(\arcsec/century)	& (km s$^{-1}$) & (km s$^{-1}$) \\ \hline
WR 48a	&	10	&	320	&	2.73	&	8.53  	& 1550   & $1900\pm200$  \\
WR 112	&	18	&	360	&	1.42	&	7.10    &  1150  &  $1230\pm260$  \\
WR 125	&	7	&	196	&	2.40	&	8.53	&  2400  &  2700 \\
WR 137	&	10	&	131	&	2.46	&	18.77   &  1700  & $1700\pm100$  \\
WR 140	&	17	&	135	&	2.67	&	39.0	& 2600   & 2860 \\ \hline    \hline
    \end{tabular}
    \caption{Measured dust rings around our sample and inferred properties with the ring spacing based on averages in multiple directions. We estimate the error on our dust-ring spacing to be $\sim$0.3\arcsec. For comparison, we show the directly measured proper motion for WR 140 as reported by \citet{Lieb2025}. We also include our estimated velocity of the dust and the terminal wind speeds (Table \ref{tab:sample}).}
    \label{tab:lifetime}
\end{table}

In addition to estimating the dust detection times for these systems, we can measure the proper motion of the dust and compare the projected velocities, which depend on the adopted distances to the stars, with the terminal velocities of the WR winds. \citet{Lieb2025} used multi-epoch imaging of WR 140 to directly measure the proper motion of the dust and found that the dust is moving out to the interstellar medium at roughly the same velocity as the stellar wind. Prior to this, \citet{2022NatAs...6.1308L} found that by using the shell separations, the distance to the star, and the orbital period, that the dust was moving at $\sim2600$ km s$^{-1}$ compared to the terminal wind speed of $v_\infty = 2860$ km s$^{-1}$ \citep{1989MNRAS.240..445W}. We carried out the same analysis for our targets, using distance estimates discussed in Section 2.
The results are tabulated in Table \ref{tab:lifetime} with the average spacing of the shells indicated in Figs.~\ref{fig:radial1} and \ref{fig:radial2}.

\section{Discussion}


\textit{JWST} has revealed large-scale dust structures around five Galactic WCd binaries, including WR 140 \citep{2022NatAs...6.1308L, Lieb2025} and the four systems presented here. In this presentation of the data taken with MIRI for WR 48a, WR 112, WR 125, and WR 137, along with the previously published data of WR 140, we can conclude that dust must be long-lived around these systems. The lack of dust seen at larger distances from the central binaries in this study is limited more by the sensitivity of the observations than by an absence of dust as the radial profiles show a decline to nearly the background level around these stars, where any dust contribution has a signal-to-noise near $\sim2$. Furthermore, the decline in the dust emission is likely a temperature effect as the temperature of an amorphous carbon grain in radiative equilibrium with a hot star at distance $r$ falls off as $T \propto r^{-0.4}$ \citet{1987A&A...182...91W}.

In all five systems, the lifetime of the dust as it approaches the interstellar medium is seen to be over 100 years. Such detection/survival times without any noticeable deceleration from ISM interactions may imply that these binaries are an important factor in the dust budget of star-forming galaxies. We note that \citet{2024ApJ...969..140R} determined that the recent dust creation episode for WR 125 created 1.5$\times 10^{-6} M_\odot$ of dust over the three-year outburst. If the same amount is created every periastron, then the MIRI observations here would imply about $10^{-5} M_\odot$ of WC dust in the image. We also note that WR 125, having one of the largest distances and only exhibiting a short time-period of dust creation near periastron, is likely the binary in our sample that produces the least amount of dust. 

Dust survivability around WCd binaries may be of cosmological importance. \citet{2020ApJ...898...74L} examined dust production from supernovae, WCd binaries, and AGB stars in stellar populations. At lower metallicity, like in the LMC, AGB stars only produce 2--3 times more dust than WCd stars, where AGB stars are often considered the biggest contributors for the dust budget, although this dust moves into the ISM with speeds two orders of magnitude slower than the WCd dust around these systems. In a population more like the Galaxy with solar metallicity, WCd binaries are the dominant source of dust until 60 Myr after star formation ceases and AGB stars begin producing dust. The long-term survivability of this dust around these systems observed here exemplifies the importance of these systems.

\begin{figure}[t]
    \centering
    \includegraphics[width=\linewidth]{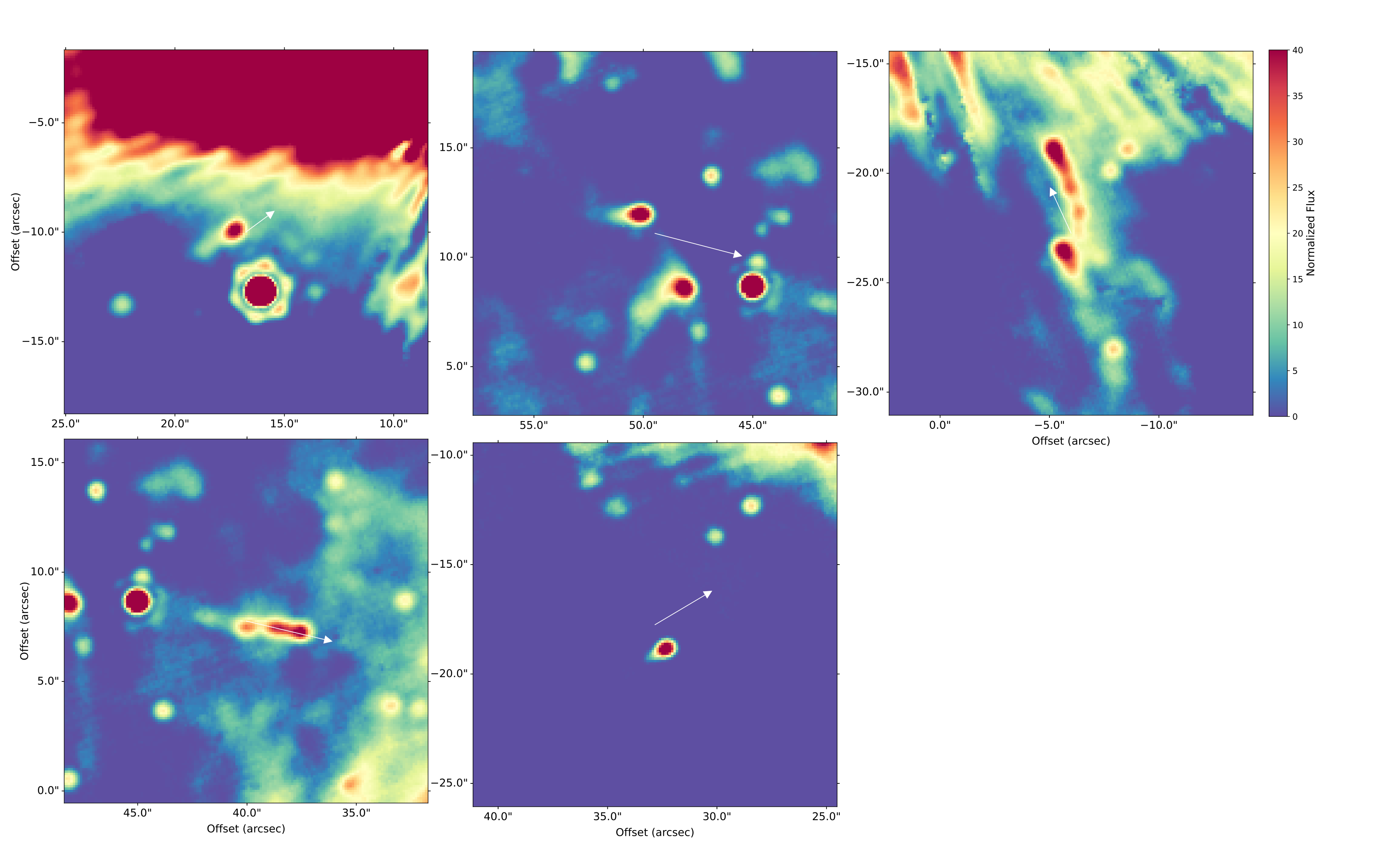}
    \caption{Potential proplyds or dusty clumps being sculpted by the radiation field from WR 48a shown in the 21$\upmu$m data. Positions are relative to WR 48a, with north being up and east to the left. The white arrows indicate the radial vector to WR 48a. The top right panel includes two potential proplyds while the other panels each include one. The separations of the proplyds are 15.5\arcsec\ (upper left), 46.5\arcsec\ (upper middle), 20.3 and 24.8\arcsec\ (upper right), 36.6\arcsec (lower left) and 33.1\arcsec\ (lower middle). We note that at the distance of this cluster, a projected separation of 1\arcsec\ is equivalent to 0.18 pc in the plane of the sky.}
    \label{fig:proplyds}
\end{figure}

We observe isomorphic concentric structures in all cases. Only one of the four systems has a well-established orbit, namely WR 137 \citep{2024ApJ...977...78R}. The only other system in our sample with a published orbit is WR 125 \citep{2024ApJ...969..140R}, which has a period based on dust outbursts. The close-in geometry of WR 112 led \citet{2020ApJ...900..190L} to infer a period for the system of $\sim20$ years. The $\sim 32$ yr period for WR 48a is only the interval between two photometrically observed maxima \citep{2012MNRAS.420.2526W}. In order to demonstrate that the events are indeed periodic, we would require further photometric monitoring covering additional maxima, which would take $\sim$a century for the timescales inferred already. The discovery of the concentric rings in this study (Figs.~\ref{fig:rgb} and \ref{fig:radial1}) extending from WR 48a demonstrates that the infrared maxima occurred regularly in the past and that the $\sim$32 year interval is indeed a periodic phenomenon tied to its binary orbit. This same logic also applies to WR 112 and WR 125. The periodic behavior of WR 137 is well established \cite[e.g.][]{2023ApJ...956..109P} and the rings confirm this here, along with the edge-on nature seen previously \citep{2024ApJ...963..127L}.

Our sample has the advantage that each target has at least an estimate of its orbital period, as shown in Table~\ref{tab:sample}. In each image, we were able to measure an average ring spacing for the concentric rings using Gaussian fits for each identified peak in Figs.~\ref{fig:radial1} and \ref{fig:radial2}. Thus, our measured average peak separation for the concentric rings, along with the distances and orbital period given in Table~\ref{tab:sample}, provide a means to estimate the proper motion of the dust shells, which was also done for WR 140 by \citet{2022NatAs...6.1308L}. WR 48a then has a proper motion of the dust shells of $\sim1600$ km s${^{-1}}$, comparable to the average terminal wind speeds of WC8 stars found by \citet{1994MNRAS.269.1082E}. WR 112's dust is propagating with a speed of $\sim$1150 km s$^{-1}$, nearly identical but not as fast as the measured terminal wind speed of 1230 km s$^{-1}$ \citep{2020ApJ...900..190L}. WR 125 has a distance of 6.58 kpc according to recent \textit{Gaia} results, implying its dust is moving with a speed of 2650 km s$^{-1}$, comparable to the terminal wind speed of the system \citep{1994MNRAS.269.1082E}. Finally, we also find that the dust around WR 137 must be moving at a comparable speed to the wind's terminal velocity, 1700 km s$^{-1}$. As WR 140 shows a similar trend in single epoch measurements of the shells and implied proper motion \citep{2022NatAs...6.1308L,Lieb2025}, we conclude that dust is propagating to the interstellar medium with speeds approximately equal to the terminal wind speed for all of our target binaries, which should be similar for all WCd binaries. This is consistent with the 3D models of stellar winds \citep{2022A&A...665A..42M} that show clumpy material moving at slower speeds than the stellar winds. Given our dust velocities have errors that likely propagate to at least $\pm100\ {\rm km\ s}^{-1}$, this likely implies the dust is moving slightly slower than the terminal wind speeds of the stars.  

Around two of our targets, namely WR 125 and WR 137, we see large diffuse components of the interstellar medium colored red in Fig.~\ref{fig:rgb}. These clouds are likely just dust in the Galactic Plane, and not necessarily near the WC binaries. If they were reflection nebulae seen near the star, they would likely appear much bluer as the stars emit more light at shorter wavelengths. The presence of these clouds do not impede our ability to detect the circumbinary dust structures around the systems. 
Our measurements and estimates of the proper motions of the dust show that the dust seems to propagate out to the interstellar medium with a velocity comparable to the terminal wind speeds of these systems. The ISM around WR 125 has been mapped in the radio and IR by \citet{1991A&A...250..171A}, who used the {\it IRAS} HCON1 images to detect a bright and extended infrared source at 60 and 100 $\upmu$m coincident with the WR star. The extended emission seen in our 21-$\upmu$m image (Figs. \ref{fig:rgb} and \ref{fig:radial2}) may be associated with the detection of the ISM by \citet{1991A&A...250..171A}. Similarly, the diffuse structure around WR 137 could be associated with Cyg OB1.


WR 48a resides in the region near the clusters Danks 1 (C 1310–624) and Danks 2 (C 1309–624) \citep{1983A&A...118..301D,1984A&A...132..301D}. The \textit{JWST} image of WR 48a shows, on close inspection (Fig.~\ref{fig:proplyds}), six objects that resemble the proplyds in the Orion nebula \citep{2008AJ....136.2136R}. Five of the six objects show evidence of photoevaporated and wind-ablated tails \citep{1996ApJ...465..216H} pointing radially away from WR 48a. Such objects are not seen in the environments surrounding the other WR targets, their appearance here may be due to the denser circumstellar environment around the Danks 1 and Danks 2 clusters in the vicinity of WR 48a. We are planning a full analysis and study of these objects including mass estimates and spectral indices (Han et al., in prep.).

The light curve of WR 48a \citep{2012MNRAS.420.2526W} shows a long-term trend that presumably peaks near periastron as is the case for WR 140 \citep{2009MNRAS.395.1749W}, WR 137 \citep{ 2023ApJ...956..109P}, and WR 125 \citep{2024ApJ...969..140R}. Superimposed on this long-term trend are small-amplitude flux increases at random times between the large maxima. These proplyd-shaped clumps could represent large clumps of dusty material ejected from the central binary during these mini-outbursts. On the other hand, they could mark sites of low-mass star formation as seen in Orion. One test of these hypotheses would be to observe WR 48a with \textit{JWST} and MIRI again. Then, if they are clumps, they would be moving away from the system where photoevaporation could help sculpt these clumps to show their current geometry. 

\section{Future Work}

This paper presents our new data on WCd binaries taken with \textit{JWST} and MIRI, with an introduction to the data and some immediate results. In our future work on these objects, we will study the temperature, dust grain sizes, and mass distributions around these stars while also examining the clumpy structures in the rings. These systems can be used to understand how the geometric structures relate to the orbital geometry. We also plan to compare the imaging to contemporaneous ground-based near-infrared imaging of all of these systems allowing us to study the evolution of the dust geometry from the inner-most to outer-most shells. Future imaging of the systems with \textit{JWST} will provide direct measurements of the proper motions of the dust and could determine whether the proplyd-like structures near WR 48a are forming stars or are low-density ejected clumps from the WR wind. Spectroscopy of the dusty structures with \textit{JWST} can constrain models of how the dust changes with time along and illuminate the evolution of the molecular precursors of the dust. Further investigations with sub-mm arrays like ALMA should be able to probe the coolest dust at larger distances than what \textit{JWST} can observe. 

\begin{acknowledgments}

This material is based upon work supported by NASA under award number 80GSFC24M0006 and based on observations made with the NASA/ESA/CSA James Webb Space Telescope. The data were obtained from the Mikulski Archive for Space Telescopes at the Space Telescope Science Institute, which is operated by the Association of Universities for Research in Astronomy, Inc., under NASA contract number NAS 5-03127 for \textit{JWST}. These observations are associated with program \#4093. Support for program \#4093 was provided by NASA through a grant from the Space Telescope Science Institute, which is operated by the Association of Universities for Research in Astronomy, Inc., under NASA contract NAS 5-03127.

N.D.R. is grateful for support from the Cottrell Scholar Award \#CS-CSA-2023-143 sponsored by the Research Corporation for Science Advancement.
E.P.L. and J.L.H. are grateful for support from a NASA FINESST fellowship under grant \#80NSSC24K1547.
C.K. acknowledges support from the Embry-Riddle Aeronautical University's Undergraduate Research Institute. 
J.R.C. acknowledges funding from the European Union via the European Research Council (ERC) grant Epaphus (project number 101166008).
M.F.C. and K.H. are supported by NASA under award number 80GSFC24M0006.
A.A.C.S. is supported by the German \emph{Deut\-sche
For\-schungs\-ge\-mein\-schaft, DFG\/} in the form of an Emmy Noether
Research Group -- Project-ID 445674056 (SA4064/1-1, PI Sander).
N.S-L. acknowledges financial support from the National Sciences and Engineering Council (NSERC) of Canada.

\end{acknowledgments}

\vspace{5mm}
\facilities{JWST(MIRI)}

\software{astropy \citep{2013A&A...558A..33A,2018AJ....156..123A}
          }

\bibliography{sample631}{}
\bibliographystyle{aasjournalv7}

\end{document}